\documentclass[preprint,amsmath,amssymb,aps]{revtex4-2}

\usepackage{graphicx}
\usepackage{dcolumn}
\usepackage{bm}
\usepackage[hidelinks]{hyperref}
\usepackage{mathtools}
\usepackage[T1]{fontenc}
\usepackage[utf8]{inputenc}

\usepackage{xspace}
\usepackage[most]{tcolorbox}
\tcbset{colback=white,colframe=black,arc=2pt,boxrule=0.5pt}
\newtcolorbox{predictionbox}[2][]{enhanced,breakable,title={#2},fonttitle=\bfseries,#1}

\usepackage{xcolor}
\newcommand{\rev}[1]{#1}

\newcommand{\ii}{\mathrm{i}}

\newcommand{\gMN}{\hat{g}_{\mu\nu}}

\newcommand{\pM}{\hat{P}_\mu}

\newcommand{\JAB}{\hat{J}_{ab}}

\newcommand{\CMMM}{\hat{C}_{\mu\nu}{}^{\rho}}

\newcommand{\Exp}[1]{\left\langle #1 \right\rangle}
\newcommand{\diag}{\mathrm{diag}}

\newcommand{\Cg}{\mathcal{C}_g}
\newcommand{\FRW}{\textsc{FRW}\xspace}

\begin{document}

\title{Quantum geometry from commutators: a Heisenberg-picture framework%
\texorpdfstring{\\}{ }%
and a toy application to early structure}

\author{Vahid Kamali}
\email{vahid.kamali2@mcgill.ca}
\affiliation{Department of Physics, McGill University, Montr\'{e}al, QC H3A 2T8, Canada}
\affiliation{Department of Physics, Bu-Ali Sina (Avicenna) University, Hamedan 65178-016016, Iran}
\affiliation{School of Continuing Studies, McGill University, Montr\'{e}al, QC H3A 2T5, Canada}
\affiliation{Trottier Space Institute, Department of Physics, McGill University, Montr\'{e}al, QC H3A 2T8, Canada}

\date{\today}

\begin{abstract}
We develop a Heisenberg-picture \emph{kinematical} framework in which (i) time is treated as a quantum observable, admitting both a relational POVM construction for semibounded spectra and a fully self-adjoint realization on an enlarged (conjugate-energy) Hilbert space enabled by a gravitational conjugation symmetry $\Cg$, and (ii) the generators of spacetime translations need not commute in curved backgrounds. The central postulate,
$[\,\hat{x}_\mu,\hat{P}_\nu\,]=\ii\hbar\,\hat g_{\mu\nu}(\hat{x})$,
makes the spacetime metric a \emph{metric operator} defined by the symmetrized commutator. Jacobi identities close the algebra and imply an operator form of metric compatibility; in a worked \FRW\ example we obtain
\rev{$[\,\hat{P}_0,\hat{P}_i\,]=2\ii\hbar\,N^2(t)\,H(t)\,\hat{P}_i$, which reduces to $2\ii\hbar\,H\,\hat{P}_i$ in cosmic-time gauge $N=1$,}
exhibiting Hubble--controlled non-commuting ``translations.'' A key structural ingredient is the symmetry $\Cg$: an antiunitary map that flips all translation generators, $\hat P_\mu\!\to\!-\Theta \hat P_\mu \Theta^{-1}$, while covariantly transforming the metric and Lorentz sectors, leaving the canonical commutators and the $[P,P]$ algebra invariant.
We discuss uncertainty relations and show how metric-operator fluctuations can rescale primordial amplitudes; an explicitly labeled \emph{toy} propagation of such a rescaling to high-$z$ halo abundances is given in Appendix~$D$.
\end{abstract}

\maketitle

\section{Introduction}

Reconciling quantum mechanics with dynamical spacetime motivates elevating geometric data to quantum operators. Here we develop a compact, explicitly \emph{kinematical} framework where the spacetime metric arises \emph{from commutators}. The guiding observation is twofold:
(i) in relativity the time interval depends on the observer, motivating time as an observable at the quantum level; (ii) in curved spacetime the naive notion that translations commute can fail---the covariant derivative algebra contains curvature and, more generally, torsion \cite{Wald:1984,MTW:1973,BirrellDavies:1982}.

Our starting point is the unified canonical relation
\begin{equation}
[\,\hat{x}_\mu,\hat{P}_\nu\,]=\ii\hbar\,\hat g_{\mu\nu}(\hat{x}),
\label{eq:xP}
\end{equation}
which defines the Hermitian \emph{metric operator} by its symmetrized form,
\begin{equation}
\hat g_{\mu\nu} \equiv \frac{1}{2\ii\hbar}\!\left(\,[\hat{x}_\mu,\hat{P}_\nu]+[\hat{x}_\nu,\hat{P}_\mu]\,\right),
\qquad
\eta_{\mu\nu}=\mathrm{diag}(-,+,+,+),
\label{eq:gdef}
\end{equation}
and reduces to Minkowski space in the flat limit. Demanding algebraic closure fixes the ``structure functions'' appearing in the commutator of translation generators and yields an operator form of metric compatibility. In a worked \FRW\ example we find
\rev{$[\,\hat{P}_0,\hat{P}_i\,]=2\ii\hbar\,N^2(t)\,H(t)\,\hat{P}_i$ (Sec.~\ref{sec:FRW}), reducing to $2\ii\hbar H \hat P_i$ in cosmic-time gauge $N=1$,}
making explicit that expansion can drive non-commuting translation generators.
\paragraph*{Main algebra and key results:}
\begin{itemize}
\item \textbf{Metric from commutator:} Eqs.~\eqref{eq:xP}--\eqref{eq:gdef}.
\item \textbf{Kinematical postulates:} Eq.~\eqref{eq:postulates}.
\item \textbf{Jacobi closure / operator compatibility:} Eq.~\eqref{eq:Cexpress}.
\item \textbf{Lorentz/tetrad extension:} Eq.~\eqref{eq:PP-Lorentz}.
\item \textbf{\FRW\ worked example:} Eq.~\eqref{eq:FRW-PP}.
\rev{\item \textbf{Weak-field check beyond \FRW:} Sec.~\ref{sec:weakfield}.}
\item \textbf{Gravitational conjugation symmetry:} Eqs.~\eqref{eq:intro-Cg} and \eqref{eq:gravC}.
\end{itemize}

\paragraph*{Gravitational conjugation (overview).}
A central structural result of this paper is a \emph{gravitational conjugation} symmetry $\Cg$:
let $\Theta$ be antiunitary ($\Theta\,\ii\,\Theta^{-1}=-\ii$); define
\begin{eqnarray}
\Cg:\;
\hat x^\mu \mapsto \Theta \hat x^\mu \Theta^{-1},\quad
\hat P_\mu \mapsto -\,\Theta \hat P_\mu \Theta^{-1},\quad\\
\nonumber
\hat g_{\mu\nu} \mapsto \Theta \hat g_{\mu\nu}\Theta^{-1},\quad
\hat J_{ab}\mapsto \Theta \hat J_{ab}\Theta^{-1}.
\label{eq:intro-Cg}
\end{eqnarray}
This map preserves the canonical commutators and the full $[P,P]$ algebra (torsion transforms odd, curvature even), hence it is a symmetry of the kinematics.
Physically, $\Cg$ organizes negative-energy solutions into a conjugate sector while preserving the operator algebra, in analogy with the Dirac reinterpretation of negative-energy solutions in relativistic quantum mechanics \cite{Dirac:1928}.
We stress that this paper is kinematical: we do not propose a full interacting field theory of sector dynamics.

\paragraph*{Time as an observable (overview).}
Treating time as an operator raises the standard Pauli-type obstruction for semibounded Hamiltonians \cite{Pauli:1933,AharonovBohm:1961,Muga:2008}. In generally covariant and/or relational settings, operational time can be implemented via POVMs and conditional probabilities \cite{PageWootters:1983,Rovelli:1991,BuschGrabowskiLahti:1995}.
Here we present both: a conservative POVM realization (semibounded spectra) and a more symmetric self-adjoint realization on an enlarged Hilbert space once $\Cg$ is included.

\paragraph*{Contributions.}
This paper: (1) states a closed operator algebra in which the metric is a commutator and Jacobi identities imply operator metric-compatibility; (2) provides two consistent realizations of time as an observable (POVM and self-adjoint on an enlarged spectrum); (3) constructs explicit representations, including a simple \FRW\ model yielding non-commuting translation generators; (4) introduces and proves invariance of the gravitational conjugation symmetry $\Cg$ that maps a sector to its conjugate while preserving the algebra; (5) gives a clearly labeled \emph{toy} phenomenological appendix (Appendix~$D$) showing how a small rescaling of primordial amplitudes can be propagated to high-$z$ halo abundances\rev{; and (6) adds a compact weak-field (linearized) check beyond \FRW\ to show how the algebra reduces at first order and how $\Cg$ acts in that limit (Sec.~\ref{sec:weakfield}).}

\paragraph*{Related work.}
Our construction is complementary to generalized uncertainty principles and minimal-length scenarios \cite{Hossenfelder:2012jw,Tawfik:2015rva}, to noncommutative spacetime/geometry frameworks (e.g.\ Snyder-type algebras and DFR-type quantum spacetime models) \cite{Snyder:1947,DoplicherFredenhagenRoberts:1995,Connes:1994}, and to current-algebra approaches to gravity and particle physics \cite{Johnson:2016smy,Johnson:2022lcj}.
The \FRW\ example is easiest to interpret in a frame/anholonomy (teleparallel) language; we therefore include an explicit teleparallel vs.\ Levi--Civita comparison in Appendix~$D$ (see also Refs.~\cite{Maluf:2013,AldrovandiPereira:2013,HayashiShirafuji:1979}).
\rev{To sharpen the distinction from neighboring approaches: unlike GUP-type deformations, we do not postulate a momentum-dependent deformation of $[x,p]$; unlike Snyder/DFR-type noncommutative spacetime, we do not postulate $[x^\mu,x^\nu]\neq 0$. Instead, we treat $[x_\mu,P_\nu]$ itself as geometric data (an operator-valued metric), and use Jacobi closure to fix the induced $[P,P]$ structure functions. Unlike teleparallel formulations that begin with a choice of connection/coframe, our starting point is the operator algebra; a teleparallel/anholonomy interpretation then emerges naturally as one convenient realization for the \FRW\ scalar representation.}

\begin{table}[t]
\caption{
Comparison of the present commutator-geometry framework with selected neighboring approaches.
All entries are schematic; see cited references for details.
\textit{Geometry / key output:} In this work, $\hat g_{\mu\nu}$ is defined from the symmetrized
$[\hat x,\hat P]$ commutator and Jacobi identities fix the structure functions $\hat C$ (operator
compatibility), with an optional curvature term $(i\hbar/2)\hat R\,\hat J$ when the Lorentz sector
is included. In GUP/Snyder/DFR/teleparallel-type approaches, the metric is typically an input
(or $[x,x]\neq 0$ is primary), rather than being defined from $[\hat x,\hat P]$.
}
\label{tab:comparison}
\begin{ruledtabular}
\begin{tabular}{lccc}
Framework & $[x,x]$ & $[x,p]$ & $[p,p]$ \\
\hline
This work
& $0$
& $i\hbar\,\hat g(x)$
& $i\hbar\,\hat C(x)\,p\;+\tfrac{i\hbar}{2}\hat R\,J$ \\
GUP
\cite{Hossenfelder:2012jw,Tawfik:2015rva}
& $0$ (typ.)
& $i\hbar(1+\beta p^2+\cdots)$
& model-dependent \\
Snyder \cite{Snyder:1947}
& $\sim i\ell^2 J$
& deformed
& deformed \\
DFR \cite{DoplicherFredenhagenRoberts:1995}
& $i\,\theta^{\mu\nu}$ (central/extra ops)
& $i\hbar\,\eta$
& $0$ \\
Teleparallel 
\cite{HayashiShirafuji:1979,AldrovandiPereira:2013,Maluf:2013}
& $0$
& tetrad directional derivative
& $[\nabla,\nabla]=T\nabla+R J$ \\
\end{tabular}
\end{ruledtabular}
\end{table}

This work is kinematical rather than dynamical. No field equations are assumed, and no claim is made that the operator algebra proposed here uniquely determines gravitational dynamics. Our results establish consistency, representation theory, and testable kinematical consequences; embedding into a full quantum field theory or semiclassical gravity framework is left for future work.

\section{Kinematical postulates and time as an observable}
\label{sec:postulates}

\subsection{Postulates}
We adopt the following operator algebra on a dense domain $\mathcal{D}$:
\begin{subequations}
\label{eq:postulates}
\begin{align}
[\,\hat{x}_\mu,\hat{x}_\nu\,]&=0, \label{eq:xx}\\
[\,\hat{x}_\mu,\hat{P}_\nu\,]&=\ii\hbar\,\gMN(\hat{x}), \label{eq:xPagain}\\
[\,\hat{P}_\mu,\hat{P}_\nu\,]&=\ii\hbar\,\CMMM(\hat{x})\,\hat{P}_\rho 
\quad\text{(minimal scalar sector)},
\label{eq:PP}
\end{align}
\end{subequations}
where $\gMN$ is symmetric and of Lorentzian signature in expectation values on physical states. The \emph{structure functions} $\CMMM$ will be fixed by the Jacobi identities below. Equation~\eqref{eq:PP} is sufficient for scalar representations; for fields with spacetime/tangent indices one can (and generally must) extend~\eqref{eq:PP} by a curvature term proportional to local Lorentz generators, see Sec.~\ref{sec:lorentz}.

\subsection{Time as an observable and Heisenberg evolution}
\label{sec:time}

Promoting $\hat{x}^0$ to an operator with
$[\,\hat{x}^0,\hat{P}_0\,]=\ii\hbar\,\hat g_{00}$
raises the familiar tension with Pauli-type arguments when $\hat{P}_0$ is semibounded and $\hat{x}^0$ is required to be self-adjoint \cite{Pauli:1933,AharonovBohm:1961,Muga:2008}. In the present framework this issue can be addressed in two complementary ways that correspond to different physical realizations.

\smallskip
\noindent\textbf{(i) Relational / POVM implementation (semibounded spectra).}
For systems whose Hamiltonian spectrum remains semibounded, $\hat{x}^0$ may be treated as a maximally symmetric operator associated with a positive-operator-valued measure (POVM) of time. In this form the commutator~\eqref{eq:xP} holds on a dense domain without demanding self-adjointness. Physical evolution is described in the Heisenberg picture with respect to an external parameter $\tau$ (for example, an affine parameter), and ``time'' predictions are \emph{relational}, conditioned on the outcomes of the clock POVM \cite{PageWootters:1983,Rovelli:1991,BuschGrabowskiLahti:1995,Muga:2008}.

\smallskip
\noindent\textbf{(ii) Conjugate-energy / self-adjoint realization (enlarged spectrum).}
Once $\Cg$ is included, it is natural to organize the kinematics on an enlarged Hilbert space
\begin{equation}
\mathcal{H}_{\rm phys} \;=\; \mathcal{H}_{+}\oplus \mathcal{H}_{-},
\label{eq:Hsplit}
\end{equation}
where the two sectors are exchanged by $\Cg$ (Sec.~\ref{sec:grav-antiparticles}) and the total energy spectrum need not be semibounded. In such a setting the spectral premise of Pauli-type no-go arguments is absent \cite{Pauli:1933,Muga:2008}, and one may realize $\hat{x}^0$ as a genuinely self-adjoint operator conjugate (in the sense of~\eqref{eq:xP}) to the appropriate energy generator on a common dense domain. In this realization one may work with generalized eigenstates
\[
\hat x^0|\tau\rangle=\tau|\tau\rangle,
\]
so that Heisenberg evolution proceeds directly in the physical time variable.

\smallskip
\noindent\textbf{Scope and failure modes.}
This paper does \emph{not} claim that every interacting theory admits such a self-adjoint realization. In particular, strong sector-mixing interactions (operators odd under $\Cg$) could spoil superselection and reintroduce stability issues. We therefore view (ii) as a kinematical possibility that becomes available once the spectrum is enlarged by $\Cg$, and leave a full dynamical implementation to future work.

\section{Algebraic closure: Jacobi identities and geometry}
\label{sec:jacobi}

The nontrivial Jacobi identity involving $(\hat{x}^\alpha,\hat{P}_\mu,\hat{P}_\nu)$ reads
\begin{equation}
[\,\hat{x}^\alpha,[\hat{P}_\mu,\hat{P}_\nu]\,]
= [\,[\hat{x}^\alpha,\hat{P}_\mu],\hat{P}_\nu\,]
+ [\,\hat{P}_\mu,[\hat{x}^\alpha,\hat{P}_\nu]\,].
\label{eq:Jacobi}
\end{equation}
Using~\eqref{eq:xPagain} and the minimal ansatz~\eqref{eq:PP} we obtain the operator identity
\begin{equation}
\CMMM\,\hat{g}_{\alpha\rho}
= \hat{g}_{\beta\nu}\,\partial_\beta \hat{g}_{\alpha\mu}
 - \hat{g}_{\beta\mu}\,\partial_\beta \hat{g}_{\alpha\nu},
\label{eq:metric-compat}
\end{equation}
where $\partial_\beta$ denotes the commutator action induced by~\eqref{eq:xPagain}:
for any function $f(\hat{x})$,
\begin{equation}
[\,f(\hat{x}),\hat{P}_\mu\,]=\ii\hbar\,\hat{g}_{\beta\mu}(\hat{x})\,\partial_\beta f(\hat{x})\,.
\label{eq:derivative}
\end{equation}
Multiplying~\eqref{eq:metric-compat} by the inverse metric $\hat{g}^{\sigma\alpha}$ yields%
\footnote{We assume $\hat g^{\mu\nu}$ exists as an (unbounded) operator inverse on a common dense domain $\mathcal{D}$ and preserves symmetry in expectation values; domain subtleties do not affect the algebraic identities stated here.}
\begin{equation}
\CMMM = \hat{g}^{\sigma\alpha}\left(
\hat{g}_{\beta\nu}\,\partial_\beta \hat{g}_{\alpha\mu}
 - \hat{g}_{\beta\mu}\,\partial_\beta \hat{g}_{\alpha\nu}
\right),
\label{eq:Cexpress}
\end{equation}
which is manifestly antisymmetric in $(\mu,\nu)$ as required.

Equation~\eqref{eq:metric-compat} is the operator analogue of (metric) compatibility: it asserts that the $\CMMM$ act as connection coefficients with respect to the \emph{directional derivatives} $\hat{g}_{\beta\mu}\partial_\beta$. In the classical limit $\hat{g}_{\mu\nu}\to g_{\mu\nu}(x)$, Eqs.~\eqref{eq:metric-compat}--\eqref{eq:Cexpress} reduce to
\begin{equation}
C_{\mu\nu}{}^\rho\,g_{\alpha\rho} =
g_{\beta\nu}\,\partial_\beta g_{\alpha\mu}
- g_{\beta\mu}\,\partial_\beta g_{\alpha\nu}.
\end{equation}

\subsection{Including local Lorentz generators}
\label{sec:lorentz}
For fields with tangent/spin indices one naturally enlarges~\eqref{eq:PP} to \footnote{
This operator relation is the direct quantum analogue of the classical identity
\begin{equation}
[\nabla_\mu,\nabla_\nu]
   = T_{\mu\nu}{}^{\rho}\,\nabla_\rho
     + \tfrac{1}{2}\,R_{\mu\nu}{}^{ab}\,J_{ab},
\label{eq:classical-commutator}
\end{equation}
familiar from differential geometry and from the spinor-in-curved-space
formalism \cite{BirrellDavies:1982,Wald:1984}.
Replacing $\nabla_\mu\!\rightarrow\!-\,\tfrac{\ii}{\hbar}\hat P_\mu$
promotes this geometric structure to an operator algebra,
\begin{equation}
[\,\hat P_\mu,\hat P_\nu\,]
   = \ii\hbar\,\hat T_{\mu\nu}{}^{\rho}\,\hat P_\rho
     + \tfrac{\ii\hbar}{2}\,\hat R_{\mu\nu}{}^{ab}\,\hat J_{ab},
\label{eq:quantum-commutator}
\end{equation}
where $\hat T$ and $\hat R$ are the operator counterparts of torsion and curvature.
}
\begin{equation}
[\,\hat{P}_\mu,\hat{P}_\nu\,]
= \ii\hbar\,\CMMM\,\hat{P}_\rho
+ \frac{\ii\hbar}{2}\,\hat{R}_{\mu\nu}{}^{ab}\,\hat{J}_{ab},
\label{eq:PP-Lorentz}
\end{equation}
where \(a,b\) are Lorentz (frame) indices.
The local-Lorentz algebra is
\begin{subequations}
\begin{align}
[\,\hat{J}_{ab},\hat{J}_{cd}\,] &=
\ii\hbar\big(
\eta_{bc}\, \hat{J}_{ad}
- \eta_{ac}\, \hat{J}_{bd}
- \eta_{bd}\, \hat{J}_{ac}
+ \eta_{ad}\, \hat{J}_{bc}\big),\\
[\,\hat{J}_{ab},\hat{P}_c\,] &= \ii\hbar\big(
\eta_{bc}\,\hat{P}_a - \eta_{ac}\,\hat{P}_b\big),\\
[\,\hat{J}_{ab},\hat{x}^\mu\,] &= 0\,.
\end{align}
\end{subequations}
The Jacobi identities then reproduce operator versions of the Bianchi identities. In the scalar representation (no tangent indices) the $\hat{J}_{ab}$ sector acts trivially and the curvature term in~\eqref{eq:PP-Lorentz} drops out.

\section{Representations}
\label{sec:reps}

\subsection{Coordinate representation (scalar sector)}
On $\mathcal{H}=L^2(\mathbb{R}^4, d^4x)$ let $(\hat{x}^\mu \psi)(x)=x^\mu\psi(x)$ and define
\begin{equation}
(\hat{P}_\mu \psi)(x)\;=\;-\ii\hbar\,g_{\mu\nu}(x)\,\partial_\nu \psi(x)\,,
\label{eq:coord-rep}
\end{equation}
with $g_{\mu\nu}(x)$ a given classical background (operator promotion in~\eqref{eq:xPagain} then means replacing $g\mapsto \hat{g}$). One checks that~\eqref{eq:coord-rep} realizes~\eqref{eq:xPagain} and gives
\(
[\,\hat{P}_\mu,\hat{P}_\nu\,]=\ii\hbar\,C_{\mu\nu}{}^\rho\,\hat{P}_\rho
\)
with $C_{\mu\nu}{}^\rho$ as in~\eqref{eq:Cexpress} (classical limit).

\subsection{Tetrads and spin connection (tensor/spinor sector)}
Introduce a tetrad $e^a{}_\mu(x)$ with $g_{\mu\nu}=\eta_{ab}e^a{}_\mu e^b{}_\nu$.
Define frame momenta and local generators by
\begin{align}
\hat{P}_a &= \hat{e}_a{}^{\mu}\,\pM, \qquad
[\,\hat{x}^\mu,\hat{P}_a\,]=\ii\hbar\,\hat{e}_a{}^\mu(\hat{x}), \\
[\,\hat{P}_a,\hat{P}_b\,] &= \ii\hbar\,\hat{T}_{ab}{}^{c}\,\hat{P}_c
+ \frac{\ii\hbar}{2}\,\hat{R}_{ab}{}^{cd}\,\JAB,
\end{align}
with $\JAB$ the local Lorentz generators acting on tangent indices, and $\hat{T},\hat{R}$ the torsion and curvature in the frame basis. This realizes~\eqref{eq:PP-Lorentz} and makes the Lorentz sector manifest.

\section{Worked example: spatially flat FRW}
\label{sec:FRW}

\rev{\paragraph*{FRW with lapse and the meaning of $\hat x^0$.}}
\rev{To make the time-choice explicit, consider spatially flat \FRW\ with a general lapse $N(t)$,}
\begin{equation}
\rev{ds^2 = -N^2(t)\,dt^2 + a^2(t)\,d\bm{x}^2,}
\qquad \eta=\diag(-,+,+,+).
\end{equation}
\rev{In the scalar coordinate representation~\eqref{eq:coord-rep}, we take the coordinate $t$ to be the time coordinate on the background manifold and represent the time operator as multiplication by that coordinate,
$(\hat x^0\psi)(t,\bm{x}) = t\,\psi(t,\bm{x})$, with $c=1$ so that $t$ has dimensions of length.}
\rev{For the worked example below we will frequently specialize to cosmic-time gauge $N(t)=1$, in which $t$ coincides with proper time along comoving worldlines; the lapse-dependent formulas make the gauge choice transparent.}

Choose the scalar representation~\eqref{eq:coord-rep} with
\begin{equation}
\rev{\hat{P}_0 = -\ii\hbar\,g_{00}\,\partial_t = \ii\hbar\,N^2(t)\,\partial_t,}\qquad
\hat{P}_i = -\ii\hbar\,g_{ij}\,\partial_j = -\ii\hbar\,a^2(t)\,\partial_i.
\end{equation}
Then
\begin{align}
\rev{[\,\hat{x}^0,\hat{P}_0\,] &= \ii\hbar\,g_{00}=-\ii\hbar\,N^2(t),}\qquad
[\,\hat{x}^i,\hat{P}_j\,] = \ii\hbar\,a^2(t)\,\delta^i{}_j,\\
[\,\hat{P}_i,\hat{P}_j\,]&=0,\qquad
\rev{[\,\hat{P}_0,\hat{P}_i\,]=2\,\ii\hbar\,N^2(t)\,H(t)\,\hat{P}_i,}
\label{eq:FRW-PP}
\end{align}
with
\(
\rev{H(t)\equiv \dot a(t)/a(t)}
\)
\rev{defined with respect to the coordinate time $t$. In cosmic-time gauge $N=1$, Eq.~\eqref{eq:FRW-PP} reduces to the simpler form $[\,\hat P_0,\hat P_i\,]=2\ii\hbar H \hat P_i$ quoted in the main text.}
Equation~\eqref{eq:FRW-PP} explicitly realizes the Second Principle: the generator of time translations fails to commute with spatial translation generators by an amount set by the expansion rate.

\rev{\paragraph*{Special epochs: $H=0$ and $H\to\infty$.}}
\rev{Because the noncommutativity is proportional to $H(t)$, two limiting regimes are worth making explicit.}
\rev{\emph{(i) Turning point / bounce:} if $H(t_\ast)=0$ at some instant $t_\ast$ (e.g.\ a turnaround between expansion and contraction), then $[\,\hat P_0,\hat P_i\,]_{t_\ast}=0$ and the algebra reduces smoothly (at that instant) to commuting translation generators in this scalar realization. If $H$ changes sign across $t_\ast$, the sign of the commutator flips continuously, with no algebraic inconsistency.}
\rev{\emph{(ii) Singular regime:} in classical FRW solutions with a big-bang-type singularity one has $|H(t)|\to\infty$ as $a(t)\to 0$. In the representation~\eqref{eq:coord-rep}, the commutator acts on a dense domain (e.g.\ smooth compact-support functions in $\bm{x}$ with suitable time regularity) as
$[\,\hat P_0,\hat P_i\,]\psi=2\ii\hbar\,N^2(t)H(t)\,\hat P_i\psi$,
which is well-defined for any finite $H(t)$. The limit $H\to\infty$ corresponds to an unbounded blow-up of the commutator coefficient and signals breakdown of the background representation (and, physically, breakdown of the classical FRW description) rather than an algebraic contradiction. We therefore interpret Eq.~\eqref{eq:FRW-PP} as applying on epochs where $N^2H$ remains finite, or after suitable regularization (e.g.\ bounce/cutoff) of the background.}

\rev{\paragraph*{Role of symmetry and global topology.}}
\rev{Homogeneity and isotropy are encoded here by the restriction $g_{ij}=a^2(t)\delta_{ij}$ and $N=N(t)$, which in particular implies $[\,\hat P_i,\hat P_j\,]=0$ for the scalar representation above. More general cosmologies (anisotropic Bianchi models or curved-$k$ FRW) would lead to additional nonvanishing structure functions via Eq.~\eqref{eq:Cexpress}.}
\rev{Global topology affects the Hilbert space and spectra but not the local commutator. For instance, replacing $\mathbb{R}^3$ by a 3-torus $T^3$ enforces periodic boundary conditions so that the spectrum of $\hat P_i$ becomes discrete; however the local relation~\eqref{eq:FRW-PP} remains unchanged as long as the local coframe/metric functions are the same.}

\noindent\emph{Interpretation and connection choice.}
For scalars with the Levi--Civita covariant derivative one has $[\nabla_\mu,\nabla_\nu]\phi=0$; in that realization, non-commutativity of ``translations'' for scalars does \emph{not} arise from curvature. The commutator~\eqref{eq:FRW-PP} instead aligns naturally with a frame/anholonomy (teleparallel/Weitzenb\"ock-type) interpretation where non-commuting translation generators reflect the anholonomy of a symmetry-adapted coframe \cite{Maluf:2013,AldrovandiPereira:2013,HayashiShirafuji:1979}. For spinors/tensors, curvature still enters through the Lorentz sector in~\eqref{eq:PP-Lorentz}

\noindent\emph{Conventions note.} \rev{For simplicity, in this conventions note we specialize to cosmic-time gauge $N=1$.} In the orthonormal frame one has
$\hat{P}_i^{(\mathrm{fr})}=-\ii\hbar a^{-1}\partial_i$ and 
$[\,\hat{P}_0^{(\mathrm{fr})},\hat{P}_i^{(\mathrm{fr})}\,]=\ii\hbar H\,\hat{P}_i^{(\mathrm{fr})}$.
Defining the lower-index spatial generator $\tilde P_i:=a\,\hat{P}_i^{(\mathrm{fr})}$ gives
$[\,\hat{P}_0^{(\mathrm{fr})},\tilde P_i\,]=2\,\ii\hbar H\,\tilde P_i$, matching Eq.~\eqref{eq:FRW-PP}.

\section{\rev{Second check beyond FRW: weak-field (linearized) background}}
\label{sec:weakfield}

\rev{To demonstrate that the commutator-geometry construction is not special to \FRW, we briefly consider a weak-field perturbation around Minkowski space in the scalar coordinate representation. Let}
\begin{equation}
\rev{g_{\mu\nu}(x)=\eta_{\mu\nu}+h_{\mu\nu}(x),\qquad |h_{\mu\nu}|\ll 1,}
\end{equation}
\rev{with indices raised/lowered by $\eta$ at leading order. Expanding Eq.~\eqref{eq:Cexpress} to first order in $h$ yields}
\begin{equation}
\rev{C_{\mu\nu}{}^{\rho}
= \partial_\nu h^{\rho}{}_{\mu}-\partial_\mu h^{\rho}{}_{\nu} + \mathcal{O}(h^2),}
\label{eq:C_linear}
\end{equation}
\rev{and hence, in the minimal scalar sector,}
\begin{equation}
\rev{[\,\hat P_\mu,\hat P_\nu\,]
= \ii\hbar\,(\partial_\nu h^{\rho}{}_{\mu}-\partial_\mu h^{\rho}{}_{\nu})\,\hat P_\rho
+ \mathcal{O}(h^2).}
\label{eq:PP_linear}
\end{equation}

\rev{\paragraph*{Newtonian limit (static weak field).}}
\rev{As a concrete specialization, take a static weak-field metric}
\begin{equation}
\rev{ds^2=-(1+2\Phi(\bm{x}))\,dt^2+(1-2\Psi(\bm{x}))\,d\bm{x}^2,}
\end{equation}
\rev{so that $h_{00}=-2\Phi$, $h_{0i}=0$, and $h_{ij}=-2\Psi\,\delta_{ij}$. If $\Phi$ and $\Psi$ are time-independent, Eq.~\eqref{eq:C_linear} gives, in particular,}
\begin{equation}
\rev{C_{0i}{}^{0}=\partial_i h^{0}{}_{0}=2\,\partial_i\Phi, \qquad
C_{0i}{}^{j}=-\partial_0 h^{j}{}_{i}=0,}
\end{equation}
\rev{and therefore}
\begin{equation}
\rev{[\,\hat P_0,\hat P_i\,]=2\,\ii\hbar\,(\partial_i\Phi)\,\hat P_0 + \mathcal{O}(h^2).}
\label{eq:PP_newton}
\end{equation}
\rev{Thus, in the weak-field (inhomogeneous) regime, the noncommutativity of translation generators is controlled by gradients of the Newtonian potential, and reduces smoothly to the commuting Minkowski algebra when $\partial_i\Phi=0$ (or $h_{\mu\nu}=0$).}

\rev{\paragraph*{Behavior under $\Cg$ at first order.}}
\rev{Since $\Cg$ acts by $\hat P_\mu\mapsto-\Theta \hat P_\mu\Theta^{-1}$ while transforming $\hat g_{\mu\nu}$ covariantly, the linearized structure functions in Eq.~\eqref{eq:C_linear} transform covariantly as well. Consequently the form of the $[P,P]$ algebra in Eq.~\eqref{eq:PP_linear} is preserved under $\Cg$ order-by-order in $h$ (for real classical $h_{\mu\nu}$ this is immediate, since $\Theta$ reduces to complex conjugation on $c$-number coefficients).}

\section{Conjugate-energy sector and gravitational conjugation}
\label{sec:grav-antiparticles}

Dirac resolved the negative-energy problem of relativistic quantum mechanics by reinterpreting negative-energy solutions as physical antiparticles \cite{Dirac:1928}. Here we propose an analogous organizational principle \emph{within our operator framework}: negative-energy gravitational solutions are grouped into a conjugate sector related to the positive-energy sector by a symmetry that preserves the algebra
\(
[\,\hat x_\mu,\hat P_\nu\,]=\ii\hbar\,\hat g_{\mu\nu}
\)
and
\(
[\,\hat P_\mu,\hat P_\nu\,]=\ii\hbar\,\hat T_{\mu\nu}{}^\rho\hat P_\rho + \tfrac{\ii\hbar}{2}\,\hat R_{\mu\nu}{}^{ab}\hat J_{ab}.
\)

\subsection{Gravitational conjugation: definition and algebraic invariance}
\label{subsec:grav-conj}

Let $\Theta$ be an antiunitary operator ($\Theta\, \ii\, \Theta^{-1}=-\ii$). Define the \emph{gravitational conjugation} map $\Cg$ by
\begin{eqnarray}
\Cg:\;
\hat x^\mu \mapsto \hat x'^{\mu}=\Theta \hat x^\mu \Theta^{-1},\quad
\hat P_\mu \mapsto \hat P'_{\mu}=-\,\Theta \hat P_\mu \Theta^{-1},\quad\\
\nonumber
\hat g_{\mu\nu} \mapsto \hat g'_{\mu\nu}= \Theta \hat g_{\mu\nu}\Theta^{-1},\quad
\hat J_{\mu\nu}\mapsto \hat J'_{\mu\nu}=\Theta \hat J_{\mu\nu}\Theta^{-1}.
\label{eq:gravC}
\end{eqnarray}

\paragraph*{Proposition (CCR invariance).}
The canonical relations are preserved:
\begin{equation}
[\,\hat x'^{\mu},\hat P'_{\nu}\,]
= \ii\hbar\,\hat g'^{\mu}{}_{\nu}.
\end{equation}
\emph{Sketch.}
$[\,\Theta \hat x^\mu \Theta^{-1},-\,\Theta \hat P_\nu \Theta^{-1}\,]
= -\,\Theta [\hat x^\mu,\hat P_\nu]\Theta^{-1}
= -\,\Theta (\ii\hbar \hat g^\mu{}_\nu)\Theta^{-1}
= \ii\hbar\,\Theta \hat g^\mu{}_\nu \Theta^{-1}.$

\paragraph*{Proposition ([P,P] invariance).}
With
\begin{equation}
\hat T'_{\mu\nu}{}^{\rho} \;\equiv\; -\,\Theta \hat T_{\mu\nu}{}^{\rho}\Theta^{-1},\qquad
\hat R'_{\mu\nu}{}^{\rho\sigma} \;\equiv\; \Theta \hat R_{\mu\nu}{}^{\rho\sigma}\Theta^{-1},
\label{eq:TR-transform}
\end{equation}
one has
\begin{equation}
[\,\hat P'_{\mu},\hat P'_{\nu}\,]
= \ii\hbar\,\hat T'_{\mu\nu}{}^{\rho}\,\hat P'_{\rho}
+ \frac{\ii\hbar}{2}\,\hat R'_{\mu\nu}{}^{\rho\sigma}\,\hat J'_{\rho\sigma}.
\end{equation}
\emph{Sketch.}
$[ -\Theta \hat P_\mu \Theta^{-1},- \Theta \hat P_\nu \Theta^{-1}]
= \Theta [\hat P_\mu,\hat P_\nu]\Theta^{-1}$ and rewrite the RHS in terms of primed tensors using~\eqref{eq:TR-transform} and~\eqref{eq:gravC}.

\subsection{Hilbert-space organization and superselection}
\label{subsec:antisector}

A convenient way to encode the conjugate sector is the direct-sum structure~\eqref{eq:Hsplit}.
Sector transitions are absent unless mediated by operators that are odd under $\Cg$; this provides a kinematical stability condition in a single-particle (non-field) setting. A full interacting quantum field theory implementation is beyond the scope of this paper.

\subsection{Relation to standard discrete symmetries}
\label{subsec:CPTremark}
The map $\Cg$ is not identified with charge conjugation, parity, or time reversal, nor with CPT. Operationally it is defined by its action on translation generators and geometric operators: it flips all $\hat P_\mu$ while covariantly transforming $\hat g_{\mu\nu}$ and the Lorentz sector in a way that preserves the commutator algebra. This is the only property needed in what follows.

\subsection{Concrete realization in FRW}
\label{subsec:FRW-anti}

For spatially flat \FRW\ (Sec.~\ref{sec:FRW}) with
\rev{$[\,\hat P_0,\hat P_i\,]=2\ii\hbar\,N^2(t)\,H(t)\,\hat P_i$,}
apply $\Cg$:
\(
\hat P'_0=-\Theta \hat P_0\Theta^{-1},\ \hat P'_i=-\Theta \hat P_i\Theta^{-1}
\Rightarrow
\rev{[\,\hat P'_0,\hat P'_i\,]=2\ii\hbar\,N^2(t)\,H(t)\,\hat P'_i}
\).
Thus the same Hubble-controlled non-commutativity appears in the conjugate sector; the sector exchange acts on energy-momentum expectation values, not on the algebra itself.

\section{Uncertainty relations and a toy cosmological remark}
\label{sec:uncertainty}

For any pair of operators the Schr\"odinger--Robertson bound gives
\begin{equation}
\Delta x^\mu\,\Delta P_\nu \;\ge\; \tfrac{1}{2}\,|\Exp{[\,\hat{x}^\mu,\hat{P}_\nu\,]}|
\;=\; \tfrac{\hbar}{2}\,|\Exp{\hat{g}^\mu{}_\nu}|.
\label{eq:uncertainty}
\end{equation}
Two observations follow.

\paragraph*{(i) Classical background.}
If $\hat{g}\to g(t,\bm{x})$ is classical, Eq.~\eqref{eq:uncertainty} becomes a \emph{coordinate-dependent} statement. In \FRW, e.g.\ $\Delta x^i\,\Delta P_j\ge (\hbar/2)\delta^i{}_j$. Thus, classical backgrounds alone do not alter physics.

\paragraph*{(ii) Metric--operator fluctuations.}
Novel effects arise when $\hat{g}$ has quantum fluctuations and does not commute with $\hat{P}$. Writing $\hat{g} = \bar g + \delta\hat{g}$ and expanding the Schr\"odinger--Robertson inequality yields
\begin{equation}
\Delta x^\mu\,\Delta P_\nu \;\gtrsim\; \tfrac{\hbar}{2}\Big(
|\bar g^\mu{}_\nu| + \mathcal{O}\big(\Exp{(\delta\hat{g})^2}^{1/2}\big)\Big),
\end{equation}
so that metric fluctuations act as \emph{multiplicative quantum noise} in the canonical normalization.

\paragraph*{Toy cosmological proxy (deferred).}
A minimal proxy is a small rescaling of primordial two-point amplitudes,
$P(k)\to(1+\epsilon)\,P(k)$, which can be propagated to high-$z$ halo abundances using standard Press--Schechter/Sheth--Tormen machinery. Since this step is \emph{model dependent} and not derived here from a full field-theoretic embedding, we quarantine it in Appendix~$D$ and label it explicitly as a toy illustration.

\section{Discussion and outlook}
We have presented a compact algebra where the metric is a commutator, time is an observable in a consistent POVM sense or (on an enlarged spectrum) as a self-adjoint operator, and the non-commutativity of translation generators in curved backgrounds follows from Jacobi identities. The \FRW\ example shows explicitly that \rev{$[\,\hat{P}_0,\hat{P}_i\,]\propto N^2H\,\hat{P}_i$}, with a natural anholonomy/teleparallel interpretation for the scalar realization, and \rev{Sec.~\ref{sec:weakfield} shows how the same algebra reduces smoothly in a weak-field (linearized) background.}

Several natural directions follow:
(i) develop the field-theory generalization so that $(\phi,\pi)$ inherit modified commutators from~\eqref{eq:xPagain}; (ii) analyze semiclassical states where $\Exp{\hat{g}_{\mu\nu}}$ satisfies Einstein equations (or their deformations) while $\delta\hat{g}$ seeds additional fluctuations; (iii) classify unitary representations of the enlarged algebra (including the local Lorentz sector) and connect to current algebra approaches~\cite{Johnson:2016smy,Johnson:2022lcj}; (iv) confront the framework with laboratory constraints (atomic clocks and interferometers) and cosmology (power spectra and high-$z$ structure), with the toy Appendix~$D$ serving only as an illustrative propagation of a proxy parameter.

\begin{acknowledgments}
This work is dedicated to the memory of Kian Pirfalak, Amirali Amini, Sepehr E Baba and to my fellow Iranians who have endured brutality for the simple act of demanding their rights. I am grateful to Mohammad Mahdi Mostafaei for his early involvement in this project. I also sincerely thank Keshav Dasgupta for carefully reading the manuscript and for insightful discussions. 
\end{acknowledgments}

\appendix

\section{Jacobi identity derivation of Eq.~\texorpdfstring{\eqref{eq:Cexpress}}{(9)}}
\label{app:Jacobi}
Starting from~\eqref{eq:Jacobi} with $[\,\hat{P}_\mu,\hat{P}_\nu\,]=\ii\hbar\,\CMMM\,\hat{P}_\rho$ and $[\,\hat{x}^\alpha,\hat{P}_\mu\,]=\ii\hbar\,\hat{g}^\alpha{}_\mu$,
\begin{align}
\ii\hbar\,\CMMM [\,\hat{x}^\alpha,\hat{P}_\rho\,]
&= [\,\ii\hbar\,\hat{g}^\alpha{}_\mu,\hat{P}_\nu\,]
 + [\,\hat{P}_\mu,\ii\hbar\,\hat{g}^\alpha{}_\nu\,] \nonumber\\
\Rightarrow\quad
-\hbar^2 \CMMM\,\hat{g}^\alpha{}_\rho
&= -\hbar^2 \Big(
\hat{g}^\beta{}_\nu \partial_\beta \hat{g}^\alpha{}_\mu
 - \hat{g}^\beta{}_\mu \partial_\beta \hat{g}^\alpha{}_\nu\Big),
\end{align}
which yields~\eqref{eq:Cexpress} after multiplication by $\hat{g}_{\sigma\alpha}$.

\section{Time operator: domains and relational interpretation}
\label{app:time}
We implement $\hat{x}^0$ as a maximally symmetric operator generating a POVM $\Pi_t$ (time-of-arrival/clock), following the operational viewpoint in Refs.~\cite{BuschGrabowskiLahti:1995,Muga:2008}. Heisenberg equations are written with respect to an external parameter $\tau$,
\(
\frac{d\hat{O}}{d\tau}=\frac{\ii}{\hbar}[\,\hat{H},\hat{O}\,],
\)
and probabilities conditioned on $\Pi_t$ yield operational time predictions. This avoids conflicts with a semibounded Hamiltonian spectrum while retaining~\eqref{eq:xPagain} on $\mathcal{D}$.

\section{Hermiticity and ordering}
\label{app:ordering}
In~\eqref{eq:coord-rep} one may symmetrize
\(
\hat{P}_\mu=-\frac{\ii\hbar}{2}\big( g_{\mu\nu}\partial_\nu + \partial_\nu g_{\mu\nu}\big)
\)
to improve Hermiticity with respect to the $L^2(\sqrt{|g|}\,d^4x)$ inner product. All statements in Secs.~\ref{sec:jacobi}--\ref{sec:FRW} remain unchanged at the level of (operator) structure functions.


\section{Phenomenology: a toy route to early structure}
\label{app:phenomenology}

This appendix sketches a \emph{toy} bridge from our kinematics to early structure formation.
The key assumption is that metric-operator fluctuations act as an (approximately) \emph{scale-independent} multiplicative rescaling of the primordial two-point amplitude:
\begin{equation}
P(k)\;\longrightarrow\; (1+\epsilon)\,P(k),\qquad \epsilon \equiv \tfrac{1}{2}\,\sigma_g^2,
\label{eq:eps_def}
\end{equation}
where $\sigma_g^2 \equiv \langle(\delta \hat g/\bar g)^2\rangle$ (evaluated in the relevant era) parametrizes the variance of fractional metric-operator fluctuations.
See Sec.~\ref{sec:uncertainty} for motivation; a fully consistent field-theory treatment is left for future work.

We use the standard linear form
\begin{equation}
P_{\rm lin}(k,z;\epsilon) \;=\; (1+\epsilon)\,D^2(z)\,P_0(k),\qquad
P_0(k)= A_s\,k^{n_s}\,T^2(k),
\end{equation}
with $D(z)$ the growth factor in flat $\Lambda$CDM (e.g.\ the Carroll--Press--Turner approximation \cite{CarrollPressTurner:1992}),
and a transfer function $T(k)$ (e.g.\ Eisenstein--Hu \cite{EisensteinHu:1998} ).
The overall amplitude $A_s$ is fixed by $\sigma_8$ at $z=0$. Cosmological parameter values used for the numerical illustration may be taken from Planck 2018 \cite{Planck2018:VI}.

As usual, the comoving mass is $M=(4\pi/3)\,\bar\rho_{m0}\,R^3$ with $\bar\rho_{m0}=\Omega_m\,\rho_{c0}$ and $\rho_{c0}=3H_0^2/(8\pi G)$. We define
\begin{equation}
\sigma(M,z;\epsilon) \;=\; D(z)\,\sigma_0(M)\,\sqrt{1+\epsilon},
\end{equation}
where $\sigma_0(M)\equiv \sigma(M,z{=}0;\epsilon{=}0)$.
Adopting the Sheth--Tormen (ST) multiplicity function \cite{ShethTormen:1999,ShethTormen:2002} (built on Press--Schechter logic \cite{PressSchechter:1974}),
\begin{align}
\nu(M,z;\epsilon) &= \frac{\delta_c}{\sigma(M,z;\epsilon)},\\
f_{\rm ST}(\nu) &= A\sqrt{\frac{2a}{\pi}}\,
\nu\,\exp\!\bigg(-\frac{a\,\nu^2}{2}\bigg)\Big[1+(a\nu^2)^{-p}\Big],
\end{align}
with $(A,a,p)=(0.3222,\,0.707,\,0.3)$ and $\delta_c\simeq1.686$, the differential mass function reads
\begin{equation}
\frac{dn}{d\ln M}(M,z;\epsilon)
= \frac{\bar\rho_{m0}}{M}\,f_{\rm ST}\!\big(\nu\big)\,
\frac{d\ln \sigma^{-1}_0}{d\ln M}.
\end{equation}
The cumulative abundance above a threshold $M_{\min}$ is
\begin{equation}
n(>M_{\min},z;\epsilon)=\int_{\ln M_{\min}}^{\ln M_{\max}}
\frac{dn}{d\ln M}(M,z;\epsilon)\,d\ln M.
\end{equation}

\begin{figure}[t]
  \centering
  \includegraphics[width=\columnwidth]{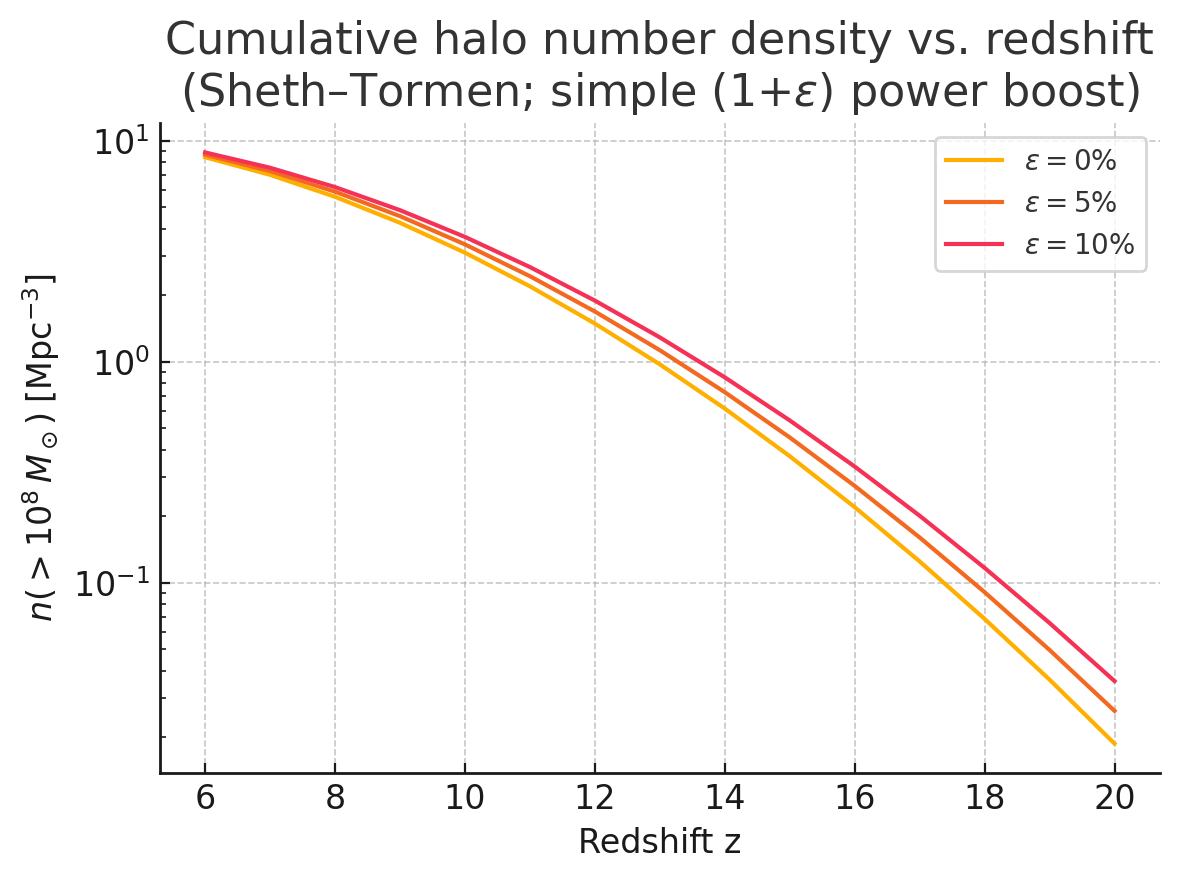}
  \caption{Comoving cumulative number density of halos above $10^8\,M_\odot$ as a function of redshift for a simple $(1+\epsilon)$ power boost, computed with the Sheth--Tormen mass function \cite{ShethTormen:1999,ShethTormen:2002} and a standard transfer function choice \cite{EisensteinHu:1998}. \textbf{Toy illustration:} the precise values depend on cosmological parameters \cite{Planck2018:VI}, transfer function details, and the redshift/scale dependence of $\epsilon$.}
  \label{fig:hmf_boost}
\end{figure}

\paragraph*{Caveats and next steps.}
(1) A rigorous embedding requires deriving how Eq.~\eqref{eq:xPagain} modifies the \emph{field} commutators $(\phi,\pi)$; here we only capture a leading normalization proxy.
(2) A scale-dependent $\epsilon(k)$ would imprint characteristic signatures in the power spectrum slope and halo mass function shape; determining $\epsilon(k,z)$ requires additional dynamics.
(3) Laboratory and CMB bounds on $\sigma_g$ must be assessed; nothing in this toy exercise guarantees viability.



\end{document}